# In vivo and in vitro study of resorbable magnesium wires for medical implants: Mg purity, surface quality, Zn alloying and polymer coating

K. Tesař[a,b,*], J. Luňáčková[c], M. Jex[d], M. Žaloudková[e], R. Vrbová[c], M. Bartoš[c,f],
P. Klein[g,h], L. Vištejnová[g,i], J. Dušková[j], E. Filová[k], , Z. Sucharda[e], M. Steinerová[k],
S. Habr[b], K. Balík[e], A. Singh[l]

[a] *Department of Materials, Faculty of Nuclear Sciences and Physical Engineering, Czech Technical University in Prague, Trojanova 13, Prague, 120 00, Czech Republic*
[b] *Institute of Physics, Czech Academy of Sciences, Na Slovance 2, Prague, 182 21, Czech Republic*
[c] *Institute of Dental Medicine, First Faculty of Medicine, Charles University and General University Hospital in Prague, Kateřinská 32, Prague, 128 01, Czech Republic*
[d] *Department of Physics, Faculty of Nuclear Sciences and Physical Engineering, Czech Technical University in Prague, Břehová 7, Prague, 115 19, Czech Republic*
[e] *Department of Composites and Carbon Materials, Institute of Rock Structure and Mechanics, Czech Academy of Sciences, V Holešovičkách 41, Prague, 182 09, Czech Republic*
[f] *Institute of Anatomy, First Faculty of Medicine, Charles University, U Nemocnice 3, Prague, 128 00, Czech Republic*
[g] *Biomedical Center, Faculty of Medicine in Pilsen, Charles University, Alej Svobody 76, Pilsen, 323 00, Czech Republic*
[h] *Department of Pathological Physiology, Faculty of Medicine in Pilsen, Charles University, Alej Svobody 76, 323 00 Pilsen, Czech Republic*
[i] *Department of Histology and Embryology, Faculty of Medicine in Pilsen, Charles University, Alej Svobody 76, 323 00 Pilsen, Czech Republic*
[j] *Institute of Pathology, First Faculty of Medicine, Charles University and General University Hospital in Prague, Studničkova 2, Prague, 128 00, Czech Republic*
[k] *Department of Biomaterials and Tissue Engineering, Institute of Physiology, Czech Academy of Sciences, Videnska 1083, Prague, 142 00, Czech Republic*
[l] *Research Center for Structural Materials, National Institute for Materials Science, 1-2-1 Sengen, Tsukuba 305-0047, Japan*

Karel Tesař [*], Corresponding author, email: Karel.Tesar@fjfi.cvut.cz
Tel: +420 721 869 628

**Abstract**

Magnesium is an excellent material in terms of biocompatibility and its corrosion products can serve as an active source for new bone formation. However, localized corrosion and $H_2$ generation limit the potential of Mg-based implants. Utilizing low-alloyed Mg-Zn wires can strongly reduce problems with large $H_2$ bubbles and improve the mechanical properties considerably while maintaining excellent long-term biocompatibility. Acidic pickling and a polymer coating can be effectively used to lower the rate of *in vivo* degradation. In this work, microstructural, mechanical, and *in vitro* characterization of 250 μm and 300 μm extruded wires made from ultra-pure Mg, commercially pure Mg, Mg-0.15Zn, Mg-0.4Zn and Mg-1Zn was performed. Additionally, Mg-0.4Zn wires together with a variant coated with a copolymer of L-lactide and ε-caprolactone were tested *in vivo* on artificially damaged Wistar rat femurs. Based on the observed Mg-induced osteogenesis, polymer-coated Mg wires with a small addition of Zn are a perspective material for bone-support applications, such as cerclage and fixation wires.

# 1 Introduction

Magnesium is one of the most promising metals for biodegradable implants. Mg and its alloys can be used for a wide range of biomedical applications [1,2]. Nevertheless, many obstacles need to be addressed during the development of a new implant. Especially when low-alloyed magnesium is used, the high corrosion rate results in hydrogen bubbles formation which can be a limitation for the healing processes [3,4]. However, the exact influence of the hydrogen bubbles *in vivo* is still a topic for debate [5]. Other detrimental factors, common for magnesium implants, include toxicity of alloying elements for some alloys [6], poor mechanical properties [7], localized corrosion [8], and the difficulty of choosing a medium to simulate the complex *in vivo* degradation as closely as possible [9]. If these obstacles were overcome, the implant in general would support bone formation [10], suppress the stress-shielding effect, common for implants from materials that have considerably higher Young's modulus than human bone [11], and degrade completely without any negative effects on the human body [12].

Thin wires are a common product and find use in various bioapplications, such as stents [13,14], bone support [15–17], and wire-reinforced composites with phosphate cement [18] or biodegradable polymer matrix [19]. Among the most important production methods for thin magnesium wires are direct extrusion [20–22] and drawing [22–24], which are both used in industrial-scale wire production. When low-alloyed magnesium is used, the onset of the degradation process needs to be controlled. This is usually performed by a coating on the implant, either by a phosphate or oxide-based layer [25] or by a layer of biodegradable polymer [26]. Since extensive deformation capability is required for the successful employment of biodegradable magnesium wires, polymer-based coatings should be of major interest. Another advantage of polymer coatings is their ability to serve as a drug delivery system [27]. Procedures like knot tying on the Mg-based sutures are a crucial part of the utilization of the wire [28]. Exceptional bending plasticity of the wires can be achieved for pure Mg by a direct extrusion resulting in beneficial microstructure and texture, allowing for mechanical twinning as a deformation mechanism in bending [29].

The aim of this work is to produce and characterize thin polymer-coated Mg wires. The exceptional bending plasticity arising from a strong extrusion texture of the wires enables load-bearing applications where large plastic deformations are needed, for instance as resorbable cerclage wires. A small addition of zinc increases corrosion resistance and prevents its localization. The effectiveness of a copolymer layer is studied to enable future functionalization. Since only biocompatible and biodegradable metals and polymers are used for the final product, there are no major concerns about hypersensitivity. This work provides a wide range of characterizations made during the development of these wires, with a final design optimized for bone support and healing. Although the preliminary investigations were carried out on the Mg-0.15Zn and Mg-0.4Zn alloys, the latter proved to be superior in every aspect. Therefore, *in vivo* tests were limited to the Mg-0.4Zn wires.

## 2 Materials and methods

### 2.1 Mg alloys, wire production and coating

Commercially pure (CP) magnesium was donated by Crown Metals CZ s.r.o. and ultra-pure magnesium was purchased from PUREMAT Technologies Sp. z o.o. The Mg-0.15Zn alloy was obtained from internal supplies and Mg-0.4Zn and Mg-1Zn alloys were produced from CP Mg and pure Zn in vacuum induction furnace Balzers VSG 02. The resulting ingots were cleaned in a 10% nitric acid and cut with electrical discharge machining (EDM) into cylindrical specimens with 6 mm diameter and 18-22 mm length. The chemical compositions of the materials are indicated in Tab. 1. In the case of ultra-pure Mg, the composition is provided by the supplier. Other values of the initial ingots were measured by glow-discharge optical emission spectroscopy (GDOES) using Spectruma GDA 750 HR, where reported values were below concentrations present in available calibration standards. To verify the concentration of Zn in the resulting wires, measurements were carried out on the Varian AA240 flame atomic absorption spectrometer (AAS).

| Element (wt. %) | Zn | Cu | Al | Si | Fe | Mg |
|---|---|---|---|---|---|---|
| ultra-pure Mg | 0.00007 | 0.000005 | 0.000005 | 0.000005 | 0.000005 | balance |
| CP Mg | 0.002 | <0.0067 | <0.045 | <0.044 | <0.0041 | balance |
| Mg-0.15Zn | 0.15 | <0.0067 | <0.045 | <0.044 | <0.0041 | balance |
| Mg-0.4Zn | 0.4 | <0.0067 | <0.045 | <0.044 | <0.0041 | balance |
| Mg-1Zn | 1.0 | <0.0067 | <0.045 | <0.044 | <0.0041 | balance |

Tab. 1. Chemical composition of materials used for thin wire production.

The 6 mm cylinders manufactured from the as-cast material (without the additional solutionizing treatment) were used to limit the number of production steps to a necessary minimum. The cylinders for the extrusion were cleaned in 10% nitric acid before the processing to remove at least 50 µm from the surface and allowed to heat up in the extrusion die for 1-2 minutes at 300 °C. Wires from CP Mg and selected alloys were produced at 300 °C via a direct extrusion method described in detail in [20], with the extrusion ratio of 1:576 and 1:400 for 250 µm and 300 µm wires, respectively. Pure Mg wires possess exceptional bending plasticity due to their texture and grain size [29]. Very good bending plasticity of the wires was maintained even for the Mg-0.15Zn and Mg-0.4Zn. An attempt was made to reliably produce wires from an Mg-1Zn alloy with a similar preparation process as of Mg-0.4Zn, however, the strengthening effect of Zn and the localization of phases were too severe to make the processing efficient in the current production setup.

The resulting wires were pickled in 7% Nital solution-based ethanol to clean the surface and prevent unwanted corrosion localization [30]. Some of the 250 µm Mg-0.4Zn wires were coated with a copolymer of L-lactide and ε-caprolactone in a 70/30 molar ratio (Purasorb

PLC 7015, Corbion). The coating medium was prepared by adding the copolymer in a 1:1 (wt.%) mixture of chloroform and acetone. The ratio of copolymer to solvent was 1:4 (wt.%). A syringe with a needle of internal diameter of 400 μm was used. Mg wire was passed through the needle and the syringe was filled with the solution. Mg wire was slowly pulled through the solution and dried for 24 hours in air at RT. The polymer coating was uniform, with a thickness of 8-15 μm.

Scanning electron microscopes (SEM) Quanta 3D FEG (FEI) and Scios 2 Dual Beam (Thermo Fisher) were used for the observation of the microstructure and focused ion beam (FIB) milling. Transmission electron microscope (TEM) JEOL JEM 2800 was used for detailed microstructural characterization and energy dispersive spectroscopy (EDS) measurements of Zn segregation. Samples for confocal laser scanning microscopy (CLSM) and light microscopy (LM) using LEXT OLS5000 (Olympus) were mechanically polished using standard metallographic methods. Where applicable, Epofix (Struers) cold-mounting resin was used. To determine the average grain size, etching in Nital was performed. The linear intercept method was used, and the resulting average intercept value was multiplied by a factor of 1.571 to consider stereological effects [31].

To assess the tensile properties, at least 3 tensile tests were performed for each material and condition on an Inspekt 100 with an initial strain rate of $10^{-3}s^{-1}$ and 30 mm gauge length. To shed light on the effect of Zn, tensile tests were performed on CP Mg, Mg-0.4Zn, and Mg-1Zn wires, according to the details in [20]. For the microhardness characterization, a Vickers microindenter Struers Duramin-2 was used for the HV0.01 measurements with a dwell time of 10 s. The average hardness value was calculated from at least six indents in the center of the longitudinal cross-sections of the wires.

**2.2 Corrosion products and degradation in artificial body fluids**

Two modifications of minimal essential medium (MEM), supplied by Thermo Fisher Scientific, were used, namely αMEM and Dulbecco's Modified Eagle Medium (DMEM). Pure αMEM and DMEM were used in the first stages of degradation experiments, including the mechanical characterization, where the degradation experiments were performed at 37 °C in an incubator without the presence of $CO_2$. For the tests addressing the morphology and composition of corrosion products of uncoated and coated wires at 37 °C, an incubator with and without 5% $CO_2$ atmosphere was used where 9 % of fetal bovine serum (FBS) and 1 % antibiotics (penicillin and streptomycin) were added to the media. The degradation tests in αMEM were conducted to observe whether the composition and purity of the material play an important role in the degradation of single wires. The times of immersion ranged from 1 hour to 2 months. The wires and the corrosion products were imaged using SEM Apreo 2S LoVac (Thermo Fisher Scientific). To determine the effect of the vacuum on the surface morphology, a digital microscope VHX-5000 (Keyence) was used as well for lower magnifications.

To determine the loss of mechanical properties of the Mg-0.4Zn wires, 60 mm long wires for mechanical testing (30 mm gauge length) were individually immersed in 45 ml of DMEM at 37 °C in air. With respect to immersion time and surface quality (Fig. 1), most wires were Nital-pickled to remove 10 μm from each point on the surface, cleaned in ethanol, dried in air, immersed in DMEM, removed after a specified time, and immediately mechanically tested. The duration of immersion was 1 day, 7 days, and 27 days, without changing the medium as for this initial study. Additionally, a set of as-extruded wires cleaned in acetone, and ethanol, dried in air, and immersed in DMEM was tested as well.

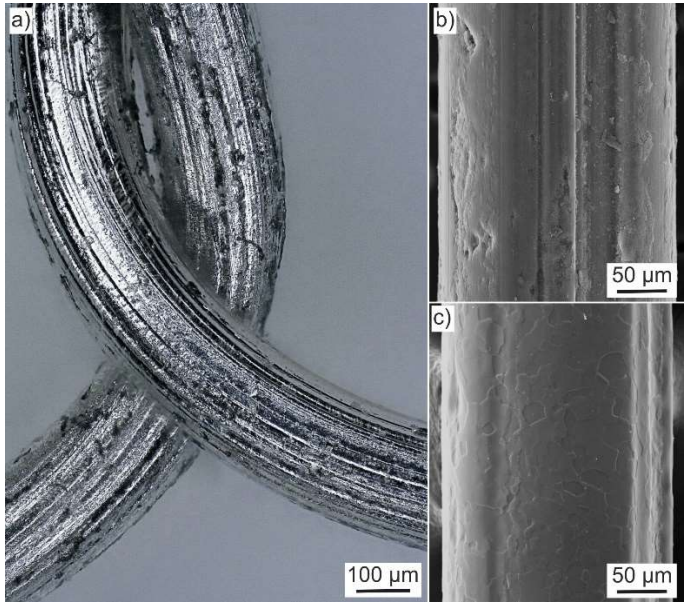

Fig. 1: a) 3D light microscope image of as-extruded 250 μm Mg-0.4Zn wire b) SEM comparison of the ultra-pure Mg cleaned in acetone and c) cleaned via Nital pickling.

### 2.3 In vitro and in vivo experiments

*In vitro*

Ultra-pure Mg wires and Mg-0.15Zn wires were dipped in 10% $HNO_3$ for several seconds, and washed in ethanol and pure water, resulting in similar surface quality as of the Nital pickling. Afterward, the wires were cut into pieces of length ranging from 1 cm to 3 cm. The MG63 human osteoblasts-like cells were seeded at the density of 77,000 cells per well in 6-well tissue culture plates in 6 ml of DMEM (Thermo Fisher Scientific, Psary, Czech Republic) + 10% of fetal bovine serum (FBS, Life Technologies, Prague, Czech Republic) + gentamicin (40 μg/ml, LEK, Ljubljana, Slovenia) and cultured in humidified atmosphere with 5% of $CO_2$. The wires were added into medium 90 minutes after seeding and the cells were cultured with them for 1, 3, and 6 days. The cells were counted on at least 10 microscopic fields for each sample using an epifluorescence Olympus IX71 microscope, obj. x10, and subsequently cell densities were calculated. Three different samples were used for each material type and length of the wire. Error bars are given as the standard error of the mean. One-way analysis of variance and the Student-Newman-Keuls Method were used for the analysis of the data (ANOVA, All Pairwise Multiple Comparison Procedures, Student-Newman-Keuls Method).

*In vivo*

In total, eleven SPF male Wistar rats (Velaz, Czech Republic) aged 4 months were used for the *in vivo* investigation using 250 μm wires. The study was divided into two parts, both lasting 1 month. In the pilot study (stage 1), a CP Mg wire, a Mg-0.4Zn wire, and a polymer-coated Mg-0.4Zn wire was implanted (one wire per animal, 3 animals in total). After the initial data evaluation, in the second stage (stage 2) of the study, two groups of four animals for each group were used for the testing of coated and uncoated Mg-0.4Zn wires. The rats were kept under conventional conditions (12/12 dark/light cycle) according to EU directive 2010/63/EU in polycarbonate 3H cages with bedding. They had free access to food and water.

The experimental surgical procedure was performed under general anesthesia. Animals were anesthetized by propofol (100 mg/kg), medetomidine (0.1 mg/kg), and nalbuphine (0.1 mg/kg) given intraperitoneally. The anesthetized animals were placed on a tempered operating table and continually supplied with oxygen via a face mask and monitored by pulse oximetry. After the cutaneous incision, access between the hind leg muscles vastus fibularis and biceps femoris was gently prepared to expose the right femur. In stage 1, two transcortical holes were drilled in the exposed diaphysis into the medullary cavity and a shallow groove was milled on the surface of the bone between the transcortical holes to induce additional damage. The test wire was bent according to the distance of drilled holes and implanted into them. In stage 2, the same implantation procedure was used except for the groove between the drilled holes as the results of the pilot study suggested that it does not offer additional data. The surgical wound was closed in three layers with an absorbable suture; the skin was treated with a liquid bandage. The animals were reverted from anesthesia with atipamezole (0.5 mg/kg) and administered with analgesics tramadol (12.5 mg/kg) and carprofen (5 mg/kg) for the next 4 days. After 4 weeks, the animals were sacrificed, and the bones were explanted and immersed in 10% formaldehyde for further analysis. The experiment was approved by the Animal Welfare Advisory Committee at the Ministry of Education, Youth and Sports of the Czech Republic (approval ID MSMT-20084/2019-3).

**X-ray microtomography**

After explantation and fixing in formaldehyde, the samples were scanned using a desktop micro-CT SkyScan 1272 (Bruker) with the following scanning parameters: pixel size of 12 μm in the first stage of the study (3 specimens) which was subsequently refined to 8 μm in the second stage (8 specimens), source voltage 80 kV, source current 125 μA, 1 mm Al filter, frame averaging (2), rotation step = 0.2° - 0.4° (depending on pixel size), rotation of 180° and scanning time 1.5-3 hours (depending on pixel size). The flat-field correction was updated before each image acquisition. The projection images were reconstructed using NRecon software (Bruker). The visualizations were acquired using DataViewer and CTVox software (Bruker). Analysis and measurements were performed in CTAn software (Bruker). The data was also subjected to semiquantitative analysis assessing the wire degradation, bone healing, reaction of the surrounding bone tissue to degradation processes and gas bubble formation.

**Histopathology**

After one week of immersion in 10% formaldehyde, the dehydration process in ethanol solutions was conducted. The bones were twice immersed in destabilized methylmethacrylate (MMA) and finally in an embedding medium (100 g MMA + 12 ml dibutyl phthalate + 1.8 g benzoylperoxide). The penetration of the medium was supported by a vacuum pump. The polymerization was made using the water bath with temperature rising by 1 °C per 2-3 days from 24 °C to 36 °C. A laboratory saw with a diamond disc was used for cutting multiple sections of each sample. The surfaces were ground on wet SiC papers and polished with 1 μm and 0.3 μm $Al_2O_3$ suspensions. The resulting 60 μm thick sections were stained for 5 minutes with a 1% solution of toluidine blue in 30% ethanol heated to 60 °C. After washing with running distilled water and differentiation in 96% ethanol, the slides were stained for 12 minutes with 0.2% solution of toluidine blue in phosphate buffer (pH = 9.1) heated to 60 °C. The slides were then washed with distilled water and dried. The sections were examined by optical microscopy using microscope Nikon Eclipse 80i, camera Jenoptik, and image analysis system NIS Elements. The bone reaction, its morphology, and the presence of fibrous tissue and cells were studied and evaluated semiquantitatively using a scoring system. Additionally, selected slides were imaged using CLSM, covered with a thin layer of carbon (8.5-9.1 nm on a Leica EM ACE600 sputter coater), and analyzed via EDS mapping using Apreo 2S LoVac (Thermo Fisher Scientific) with EDS detector Octane Elite SDDs (Ametek) at magnifications 200 – 1100x and accelerating voltage 15 kV.

## 3 Results

### 3.1 Microstructure and mechanical properties of Mg and Mg-Zn wires

The overview of the grain size and tensile mechanical properties of thin Mg wires is presented in Tab. 2. The data for the unalloyed wires are comparable to the data measured for the references [20,29], where the small differences can be attributed to the slightly altered extrusion setup. As can be seen, the Zn addition decreases the resulting grain size from 21 μm to 6 μm, i.e., for the CP Mg and Mg-1Zn wires, respectively. Furthermore, a similar level of grain refinement can be obtained by a smaller addition of Zn (0.4 wt.%). The microstructure of all wires consists of equiaxed grains. Fig. 2a shows a view of grain structure near the surface of 300 μm wire using the TEM-STEM technique. The sample was sliced perpendicular to the length of the wire near its surface by FIB technique. All grain boundaries are observed to be planar. Grain boundary junctions are triple point junctions with nearly equal angles, suggesting a near-equilibrium grain structure resulting from recrystallization. Moreover, the grain boundaries are often on close-packed planes, as shown in Fig. 2b,c. For the grain marked 1, oriented along a $\langle 1\bar{2}10 \rangle$ zone axis (Fig. 2b), the grain boundary with grain 2 is observed to be on prismatic $\{10\bar{1}0\}$ plane. Grain 3, when tilted into a $\langle 1\bar{2}10 \rangle$ zone axis (Fig. 2c), shows that its grain boundary with grain 1 is also on a prismatic $\{10\bar{1}0\}$ plane (when no Moire fringes are observed, the boundary is assumed to be parallel to the direction of view). Similarly, more examples of grains were observed where planar grain boundaries were on basal, prismatic or pyramidal planes. The grain boundaries were often segregated with the Zn element, Fig. 2d-f. Grain boundaries of grain 1 with grains 2 and 3 were found to have a Zn concentration of about 1 at.%, as compared to the matrix value of ~0.3 at.% (Fig. 2f).

| wire | $d$ (µm) | $\sigma_{02}$ (MPa) | $\sigma_U$ (MPa) | $\sigma_L$ (MPa) | $\sigma_T$ (MPa) | $A_t$ (%) |
|---|---|---|---|---|---|---|
| CP Mg, 250 µm | 21±5 | 104±4 | - | - | 213±1 | 17±1 |
| Mg-0.4Zn, 250 µm | 7±1 | 216±1 | - | - | 263±2 | 16±4 |
| Mg-0.4Zn, 300 µm | 6±1 | 223±6 | 234±8 | 224±5 | 294±7 | 22±2 |
| Mg-1Zn, 250 µm | 6±1 | 176±18 | - | - | 233±12 | 12±2 |

Tab. 2 Material characteristics of Mg and Mg-Zn wires. $d$ – average grain size, $\sigma_{02}$ – 0.2% offset yield strength, $\sigma_U$ – upper yield strength, $\sigma_L$ – lower yield strength, $\sigma_T$ – maximum true yield stress, $A_t$ – total elongation to failure

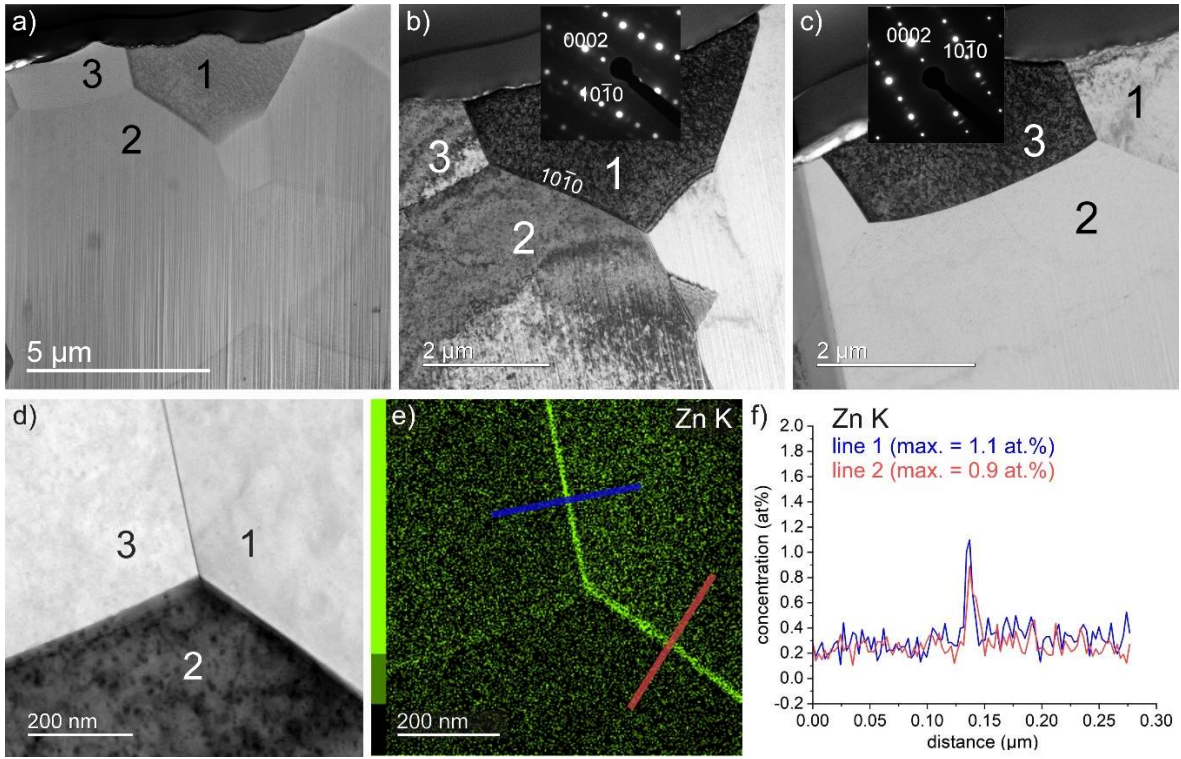

Fig. 2: Grains and grain boundaries in 300 µm Mg-0.4Zn wire observed by TEM. a) STEM image showing grains (vertical lines near the bottom are a FIB thinning artifact). b) Bright field (BF) micrograph from grain marked 1 tilted into a $\langle 1\bar{2}10 \rangle$ zone axis (diffraction pattern as an inset). c) BF micrograph from grain marked 3 tilted into a $\langle 1\bar{2}10 \rangle$ zone axis (diffraction pattern as an inset). d) STEM image of the grain boundary junction between grains marked 1, 2 and 3, and e) its respective EDS Zn map. Grain boundaries with grain marked 1 are bright in contrast due to the segregation of Zn. f) Composition line profiles along lines (blue and red) in e) are shown in the concentration plot in f).

No Mg-Zn phases were detected by means of electron microscopy for Mg and Mg-0.4Zn wires. Occasionally, a MgO particle or particle clusters were observed in all wires by the means of SEM which is related to the initial material purity [20]. The production of Mg-1Zn wire at 300°C was challenging probably due to the Mg-Zn phases which were observed in the initial ingot (as the solutionizing step is intentionally omitted from the production setup), resulting in inhomogeneous wire and plunger load during extrusion. This can be improved by solutionizing the ingot, however, the solid solution strengthening proved to be too high for the limits of the current extrusion setup.

Both yield stress and maximum true yield stress for thin Mg-based wires are substantially improved by Zn addition. The effectiveness of low-alloyed Mg can be seen in Tab. 2. The inhomogeneous Mg-1Zn wires, produced in the frame of this study, are characterized by much lower ductility and large deviations of strength parameters, resulting from the intermetallic phases figuring as prominent stress-concentrators concerning the small wire diameter and the inhomogeneous nature of the extrusion process. In this regard, lower Zn alloying provides superior properties when no solutionizing step is added to the production process. Due to their inhomogeneity and the difficulties of preparation, Mg-1Zn wires are omitted from both *in vitro* and *in vivo* characterization in this study.

Two wire diameters were produced from Mg-0.4Zn alloy: 250 μm and 300 μm. The extrusion ratios for the two diameters were 1:576 and 1:400, respectively. The main difference between the two diameters is the yield behavior. For the 250 μm wires, the yielding (with a few exceptions) was realized monotonously, resulting in the determination of 0.2% offset yield strength values, similarly as in the case of the Mg and Mg-1Zn wires. For all the 300 μm wires, distinctive upper and lower yield points emerged with an average difference between them of 10±2 MPa. These discontinuous yielding phenomena triggered by Zn are described in the literature for more complex Mg alloys and are attributed to the low grain size in the micrometer range and Zn segregation on grain boundaries [32–34]. Although the upper yield strength provides high values, for the design of biomedical implants it is safer to consider lower yield strength as a yield stress estimate. The value of the lower yield point of 224±5 MPa corresponds well to the value of 0.2% offset yield strength 223±6 MPa. The absence of a more pronounced and sharp yield point suggests that the effect of Zn segregation is limited to some boundaries, as discussed further in Section 4.1. The higher total elongation to failure is expected for the larger diameter of the wire, as the fracture is controlled by the largest of the MgO particles [20], and the MgO/Mg area ratio in each cross-section should be statistically lower for thicker wires when the same initial ingot material is considered.

As for the microhardness, 250 μm diameter wires were measured, offering a direct comparison among various materials. The microhardness value of CP Mg wire HV0.01 32±2 increased to HV0.01 45±3 and HV0.0.1 48±2 for the Mg-0.4Zn and Mg-1Zn, respectively. Therefore, the effect of the Zn solid solution strengthening is apparent for the wires.

## 3.2 Degradation in minimal essential media and the loss of mechanical properties

Typical morphologies of the 250 μm wires, without and with Nital-based acidic pickling [30] and subsequent immersion in αMEM+FBS for 1 h and 24 h, are shown in Fig. 3. Surfaces depicted by SEM illustrate the most common appearance of the wires. The corrosion behavior in the first 24 hours depends on material purity and Zn content. It is apparent from Fig. 3 that the nucleation of localized corrosion products is to some extent universal to all wire states. The residuals of the extrusion lubricants, surface damage, microcracks, Fe-rich residuals from the extrusion die, etc., provide nucleation sites for accelerated corrosion, especially for the as-extruded wires.

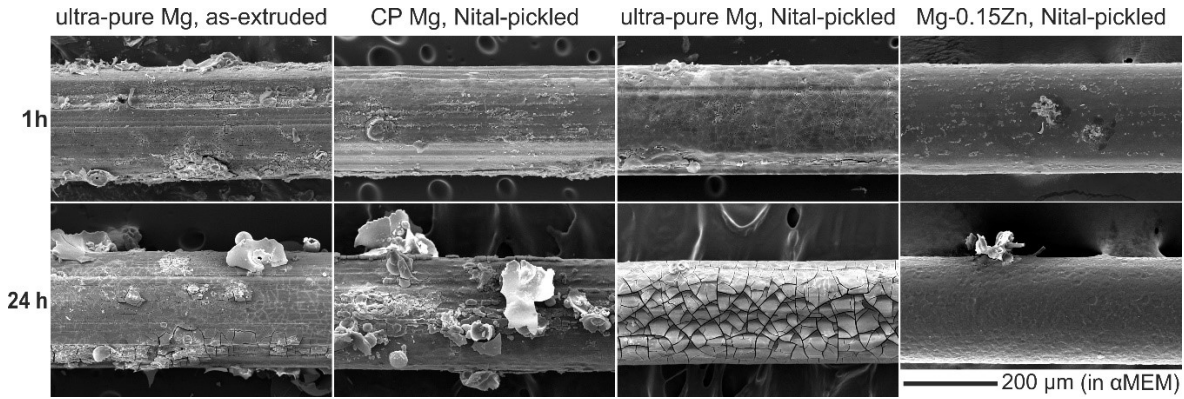

Fig. 3: Typical SEM micrographs of the degradation morphology of uncoated thin Mg and Mg-Zn wires in αMEM+FBS after 1 and 24 hours.

After 24 hours, the CP Mg and as-extruded ultra-pure Mg show many sites with morphology that relate to $H_2$ generation. This leads to the formation of (hemi)spherical or tubular features, created by the nucleation of corrosion products directly on the αMEM/$H_2$ interface [3]. These fully or partially closed thin features are shaped by the expansion of $H_2$ bubbles and the growth of corrosion products. In contrast, the Nital-pickled ultra-pure Mg shows only limited corrosion and re-deposition in the vicinity of the local corrosion sites, with lower occurrence along the wire length. The effect of re-deposition and subsequent nucleation is well-known, discussed in [35], and its mechanisms are explained in detail in [36]. An example of the redeposition, where fine corrosion products nucleate in the vicinity of a local corrosion site, can be observed in Fig. 4a.

Although the ultra-pure magnesium used in this work has very low Fe content (Tab. 1), it is apparent that the initial material purity does not offer complete protection from localized corrosion. An example of fully closed and open (hemi)spherical features that appeared after 24 h on ultra-pure Mg is presented in Fig. 4b. It is necessary to mention that the presence of these features is undesirable for implant applications due to the encapsulation of volumes making them inaccessible for cell proliferation. However, when considering the scale of the thin Mg wires, the adverse effects should be limited in the non-static conditions *in vivo* [37]. For the ultra-pure Mg, most of the wire after 24 h (Fig. 3) is composed of the typical plate-like morphology with occasional localized corrosion and $H_2$ bubble formation. This morphology is well described in the literature and the cracking of the corrosion layer is attributed to the dehydration via high vacuum in the SEM chamber [38]. This fact is supported by the 3D light microscopy images of the ultra-pure Mg wire surfaces, which do not show any cracking (Fig. 4c-e). Moreover, 3D light microscopy enables the differentiation of transparent and nontransparent/metallic areas on the wire (Fig. 4c).

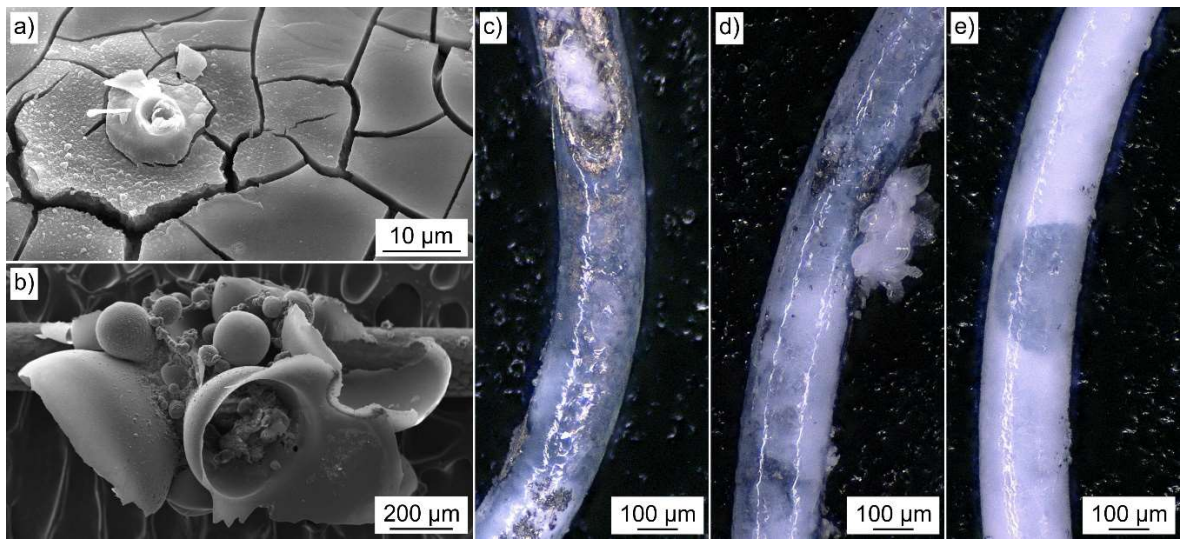

Fig. 4: Corrosion of ultra-pure Mg wire in αMEM+FBS. a) Redeposition-stimulated nucleation of fine corrosion products after 2 hours near a corrosion site, b) open and closed corrosion products formed on $H_2$ bubble cluster after 24 h, c-e) 3D light microscopy images of the wire after 1 h, 24 h and 15 days.

The Mg-0.15Zn wire after 24 h (Fig. 3) shows an area of a compact layer of corrosion products with localization of corrosion limited to one site. The compactness of the layer for some samples is probably connected with its superior mechanical properties or lower thickness since only microcracks were introduced to the layer by SEM vacuum and no major cracking was observed for some wires. However, approximately half of the Mg-0.15Zn samples also showed cracked plate-like morphology with plate size larger than in the case of ultra-pure Mg. Longer corrosion tests, for up to two months, were conducted in this study. In contrast with the first 24 hours, it is rather difficult to rigorously compare the morphologies since the rate of degradation can be highly localized. In general, no CP Mg or any wire in the as-extruded or acetone-cleaned conditions was able to stay compact after 1 month. On the other hand, most of the Nital-pickled ultra-pure Mg, Mg-0.15Zn wires (and Mg-0.4Zn wires tested in later stages of the experiments in DMEM) remained compact. However, the metal in the wires is often replaced by the corrosion products in a large volume (Fig. 4e). Based on the results of the initial degradation tests in αMEM, Zn addition was chosen as a more efficient and cost-effective method for improvement of thin Mg wire corrosion properties when compared to ultra-pure Mg wire.

To evaluate the deterioration of tensile properties after immersion in MEM, 300 μm Mg-0.4Zn wires were chosen as the most promising candidates for medical devices. The wires were Nital-pickled, reducing the wire diameter to 280 µm to obtain a defect-free surface. The initial mechanical properties of such wires agree with the values of 300 µm wires presented in Tab. 2. DMEM medium at 37 °C was selected to comply with the *in vitro* tests in this study and make an initial approximation of the *in vivo* conditions. In the frame of this study, many important controls are absent for the tensile testing of the wires, such as dynamic loading, constant change of the medium, and oxygen/$CO_2$ control [39]. These parameters will be addressed in future studies on the wires. As the degradation of mechanical properties is a complex process, the corrosion is site-specific along the wire length and the corrosion products can occupy large areas of the wire cross-sections, it may be misleading to address the properties in terms of mechanical stress. For this reason, the values reported in Tab. 3 are in terms of yield force ($F_{02}$) and maximum engineering force ($F_M$) for the resulting wire geometry, composition, and surface condition.

| DMEM immersion time (days) | $F_{02}$ (N) | $F_M$ (N) | $A_t$ (%) |
|---|---|---|---|
| 0 | 13.7±0.4 | 15.2±0.2 | 22.0±2.3 |
| 1 | 12.9±0.4 | 14.2±0.1 | 9.8±2.4 |
| 7 | 12.9±1.2 | 13.3±0.8 | 3.6±1.2 |
| 27 | 12.6±0.2 | 13.0±0.3 | 2.5±1.3 |

Tab. 3 Tensile performance of Nital-pickled Mg-0.4Zn wires (280 µm diameter) after immersion in DMEM at 37 °C. $F_{02}$ – 0.2% offset yield force, $F_M$ – maximum engineering yield force, $A_t$ – total elongation to failure.

Tab. 3 shows that the decrease of tensile strength in terms of the force is of a moderate character. The trend suggests that the loss of strength is prominent in the early stages of the degradation and then it stabilizes. Even after 27 days, the strength decrease (~8% for yield force and ~15% for maximum force) seems to be acceptable for most bioapplications that could be considered for the initial wires. However, the parameter that needs to be controlled in this case is the total elongation to failure or ductility. After 24 hours of immersion, the total elongation to failure drops from 22.0±2.3 % to 9.8±2.4 % and after 7 days it reaches only 3.6±1.2 % which is inadequate for implant applications. In accordance with the conclusions of [40], the localization of corrosion products/pits is the leading factor for the deterioration of mechanical properties. In this regard, a fragment of a wire that was tested after 27 days of immersion in DMEM and had the lowest measured properties of $F_{02}$ = 12.6 N, $F_M$ = 12.9 N, and $A_t$ = 1.4 % was tested again using the available remaining 12 mm gauge length and otherwise the same test parameters. The measured properties were thereafter $F_{02}$ = 13.2 N, $F_M$ = 13.8 N, and $A_t$ = 3.6 %. This behavior suggests that the properties of the wires are driven by the local weak spots that arise from localized corrosion phenomena. The testing of the as-extruded (acetone-cleaned) wires was not possible since the extrusion-related surface artifacts and contamination resulted in the segmentation of all acetone-cleaned 60 mm wires within the first 24 hours in at least 1 spot along the wire length. The fact that after Nital pickling the wires remain compact clearly shows the benefit of the acidic pickling procedure [30]. What remains to be addressed for biomedical applications is the corrosion-driven decrease of elongation to failure of the Nital-pickled wires upon immersion. Protective layers [26,41] and wire-stranding or braiding [23] appear to be suitable approaches to improve this parameter.

### 3.3 Mg and Mg-0.15Zn in vitro

To shed light on the effect of a small Zn alloying of Mg-based wires *in vitro*, samples were tested for their biocompatibility with MG63 human osteoblasts-like cells. Since the literature suggests that even Mg–6Zn alloy is safe for cellular applications, with a cytotoxicity grade of 0–1 [42], we only investigated the differences between ultra-pure Mg and Mg-0.15Zn. This is done mainly to assess whether the approach of alloying with Zn is more advantageous than Mg purification also for the *in vitro* tests and whether the material volume to DMEM volume plays an important role in the resulting cell densities (Fig. 5).

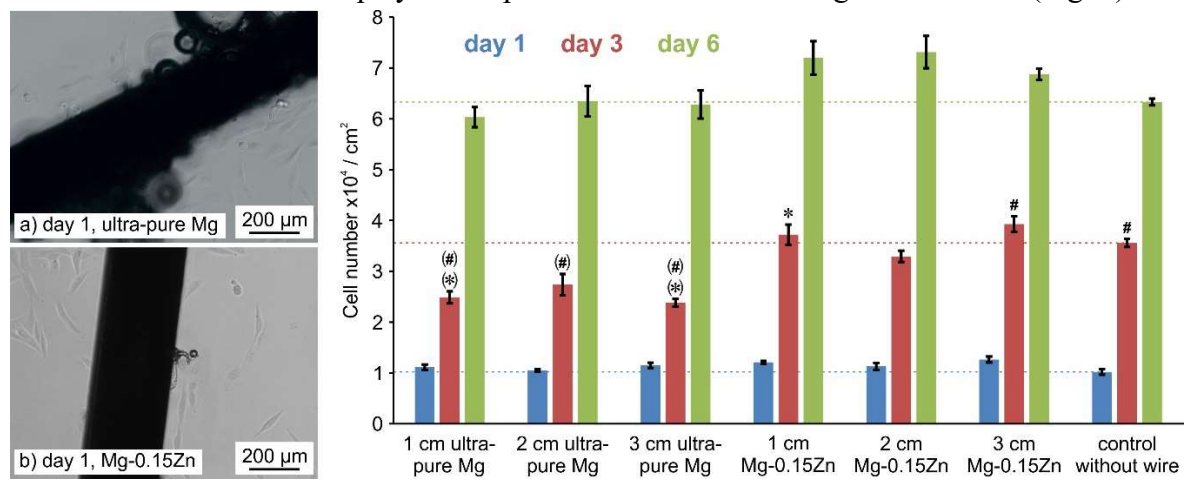

Fig. 5 Light micrographs from *in vitro* tests after 1 day; a) ultra-pure Mg, b) Mg-0.15Zn alloy 250 μm wires, c) number of MG63 human osteoblasts-like cells for 1, 3, and 6 days with respect to wire length and composition in 6 ml of medium. * and # labels statistically significant difference in comparison with (*) and (#), respectively.

Cell adhesion was similar for all samples on day 1 after seeding. The MG63 cells were spread and elongated and grew also in the vicinity of the wires. Changes in the surface morphology of the ultra-pure wires were apparent within the first day, as illustrated in the example in Fig. 5a. Subsequent hydrogen bubble formation is similar to the features in Fig. 4b and the reference [3]. Mg-0.15Zn wires appeared unchanged during the experiment (Fig. 5b). This is in accordance with the results of corrosion tests in αMEM shown in Fig. 3. Overview of the cell numbers is shown in Fig. 5c. Statistical evaluation did not provide any significant changes in cell number after the day 1. On day 3, Mg-0.15Zn wires supported cell proliferation when compared to ultra-pure Mg wires. On day 6, cell culture was confluent on all samples and the differences among samples were statistically not significant. However, the data suggest that Mg-0.15Zn is superior to ultra-pure Mg, possibly due to lower hydrogen production and the positive effects of Zn on osteoblast proliferation [43].

In general, a slower dissolution rate and lower hydrogen evolution for Mg-0.15Zn seem to be favorable for the growth of MG63 cells on the third day. On day 6, all materials supported cell growth *in vitro*. Furthermore, there was no statistically relevant observation, that the ratio of wire length to the medium volume strongly influences the outcomes of the *in vitro* tests. It is therefore reasonable to advance to the *in vivo* experiments using the Zn addition approach, which is superior regarding the initial costs of the material and the achievable mechanical properties.

### 3.4 In vivo tests on artificial bone defects

Specimens from the sacrificed animals implanted with 250 µm wires were analyzed using 2D and 3D µCT visualizations to evaluate wire degradation and neighboring bone structure. Fig. 6 shows the transaxial matched images from µCT and histological slides.

Only the transcortical hole sites and the medullary cavity were analyzed in detail by µCT, as the direct contact with the damaged cortical bone and the presence of the Mg-based wire in the medullary cavity provide the most important information concerning the possible utilization of thin resorbable wires for implant applications (e.g. cerclage wires for metacarpal fracture fixation [44] and osteotomies in hand reconstructive surgery, especially considering children and infants). To provide a comparison of the µCT data on this specific bone damage, a tailored semi-quantitative scoring system was defined, less detailed than the more complex RUST system that is recommended for tibial fractures [45] and used successfully for rat femoral fractures in [46]. In Tab. 4a, the data providing the most evident µCT outputs are summarized, focusing on the more controlled *in vivo* stage 2. The complete table, as well as the definition of the semi-quantitative scoring system used, is presented in Supplement 1.

The wire was almost completely degraded in all but one test animal (polymer-coated Mg-0.4Zn, stage 1 of *in vivo* tests). All uncoated wires – e.g. CP Mg (Fig. 6a) and Mg-0.4Zn (Fig. 6c) were degraded after 1 month of healing. The Mg-0.4Zn wire coated with biodegradable copolymer in stage 1 of the study was present in the sample along its full original length. The wire that remained was reduced to 189±20 µm in diameter (Fig. 6b). The stage 2 polymer-coated Mg-0.4Zn samples (4) were almost completely degraded. Since the single wire that remained was from the initial stage 1 of the study, where the Nital-pickling procedure was performed by a different technician, this discrepancy in the *in vivo* degradation rate vaguely suggests that the surface quality could play a significant role in the observed *in vivo* degradation rate as well (as discussed in Section 4.2, see Fig. 6b,d). However, for the determination of the actual *in vivo* degradation rate of these wires, a more complex *in vivo* study will be conducted, as this one is focused primarily on the effect of Mg wire, material composition, and polymer coating on osteogenesis and related phenomena.

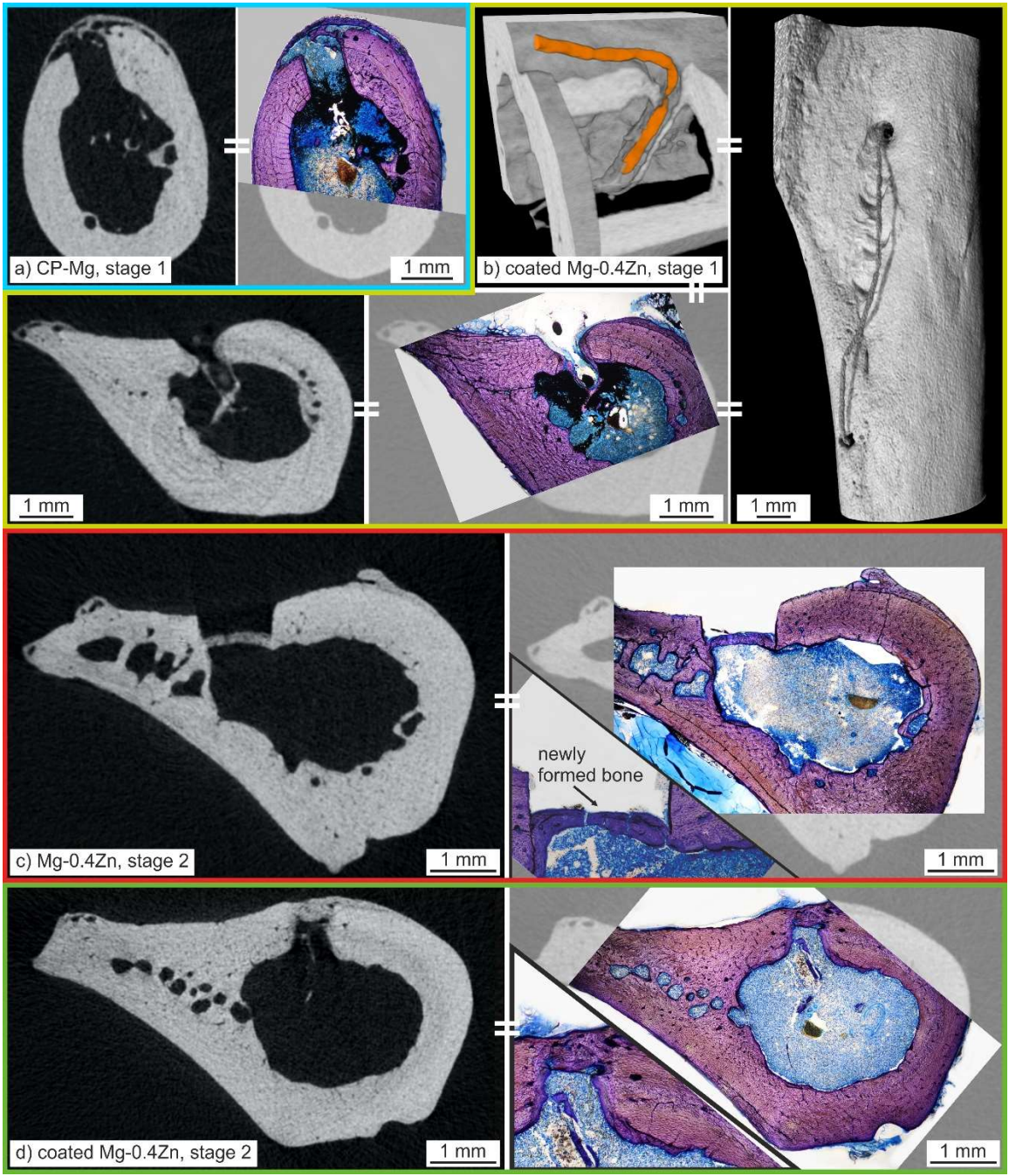

Fig. 6 Matched µCT and histological slides of rat femur (transcortical hole sites, originally with inserted wire) after 1 month of healing, a) CP Mg (distal defect), b) polymer-coated Mg-0.4Zn stage 1 including the 3D µCT reconstructions of the whole test site, detail showing remaining wire in orange color (proximal defect), c) uncoated Mg-0.4Zn stage 2 (proximal defect), d) polymer-coated Mg-0.4Zn stage 2 (proximal defect).

Regarding the semiquantitative µCT data evaluation of the *in vivo* tests (Tab. 4, Supplement 1), it seems that the PLA/PCL copolymer coating affected the *in vivo* degradation rate while no negative effects of the polymer coating on the healing processes were observed. Small particles of either residual magnesium or corrosion products could be observed for most of the polymer-coated wires and none for the uncoated wires. This is also true for intramedullary osteogenesis, where promoted osteogenesis was observed for polymer-coated wires (Fig. 7). Rather than the effect of the PLA/PCL copolymer coating itself, this relates to the slower degradation rate induced by a protective copolymer layer (yet for such thin wires and the wire surface quality reached, the degradation is almost complete). Drilled transcortical holes were not completely healed in any specimen and were covered by a layer of new bone with various thicknesses and level of continuity, where differences can be observed with respect to the stage of the wire degradation (Fig. 6). However, this should be studied in detail in subsequent *in vivo* tests.

| a) µCT | wire residuals | intramedullary osteogenesis | transcortical hole site healed |
|---|---|---|---|
| uncoated (4x) | none | limited | partially |
| polymer-coated (4x) | small particles | promoted | partially |

| b) histopathology | giant cells | macrophages | (periosteal, endosteal) remodeling |
|---|---|---|---|
| uncoated (4x) | 2 of 4 animals<br>2 of 17 slides | 3 of 4 animals<br>5 of 17 slides | (4,4) of 4 animals<br>(13,9) of (16,12) slides |
| polymer-coated (4x) | 0 of 4 animals<br>0 of 15 slides | 1 of 4 animals<br>1 of 15 slides | (4,2) of 4 animals<br>(9,4) of (15,10) slides |

Tab. 4 Major outcomes of blind semiquantitative evaluation on a series of cuts in the *in vivo* stage 2 a) µCT data in the vicinity of the transcortical hole sites, b) histopathological evaluation summarizing the results according to [47]. For a more in-depth evaluation of individual test animals and samples, see Supplements 1 and 2.

Promoted formation of new bone can be observed in the vicinity of original wire positions in the intramedullary cavity. This is further demonstrated in Fig. 7. Fig. 7a shows a coronal µCT cut through an entire scanned section of the femur implanted with polymer-coated Mg-0.4Zn wire (stage 2) that is degraded (the same test animal as for Fig. 6d). Fig. 7b,c show sagittal cut and 3D reconstruction, respectively. Fig. 7b shows, how the wire positions correspond with the osteogenesis induced in the intramedullary cavity (which is discussed in greater detail in Section 4.4.). From Fig. 7b is apparent that one end of the implanted wire was misaligned with the transcortical defect axis. This phenomenon was rare for the studied samples, as it was caused by an imperfect implantation. However, the shape of the transcortical hole (and therefore the axis of the milling) and the misaligned position of the newly formed bone in the intramedullary cavity further support the presumption that the promoted osteogenesis is indeed caused by the Mg implant. Fig. 7c shows the off-axis cut in the proximal defect (see Fig. 6d for the in-axis cut). Fig. 7e shows a ring of newly formed bone around the wire near the proximal defect that is substantially larger in diameter (> 480 µm) than the initial wire, including the polymer layer. As the wire (including the corrosion products) should be after 1 month *in vivo* significantly thinner, this further supports the idea of the preferential bone formation on the tissue/$H_2$ interface which is in accordance with the comments on the preferred nucleation of calcium phosphates *in vitro* on the MEM/$H_2$ interface [3]. Yet, the large bubbles (Fig. 7a) often do not show this encapsulation, suggesting that this should be connected either with the presence of Mg-based implant (promoting local bone formation), or by the implantation-related damage to the surrounding tissue itself.

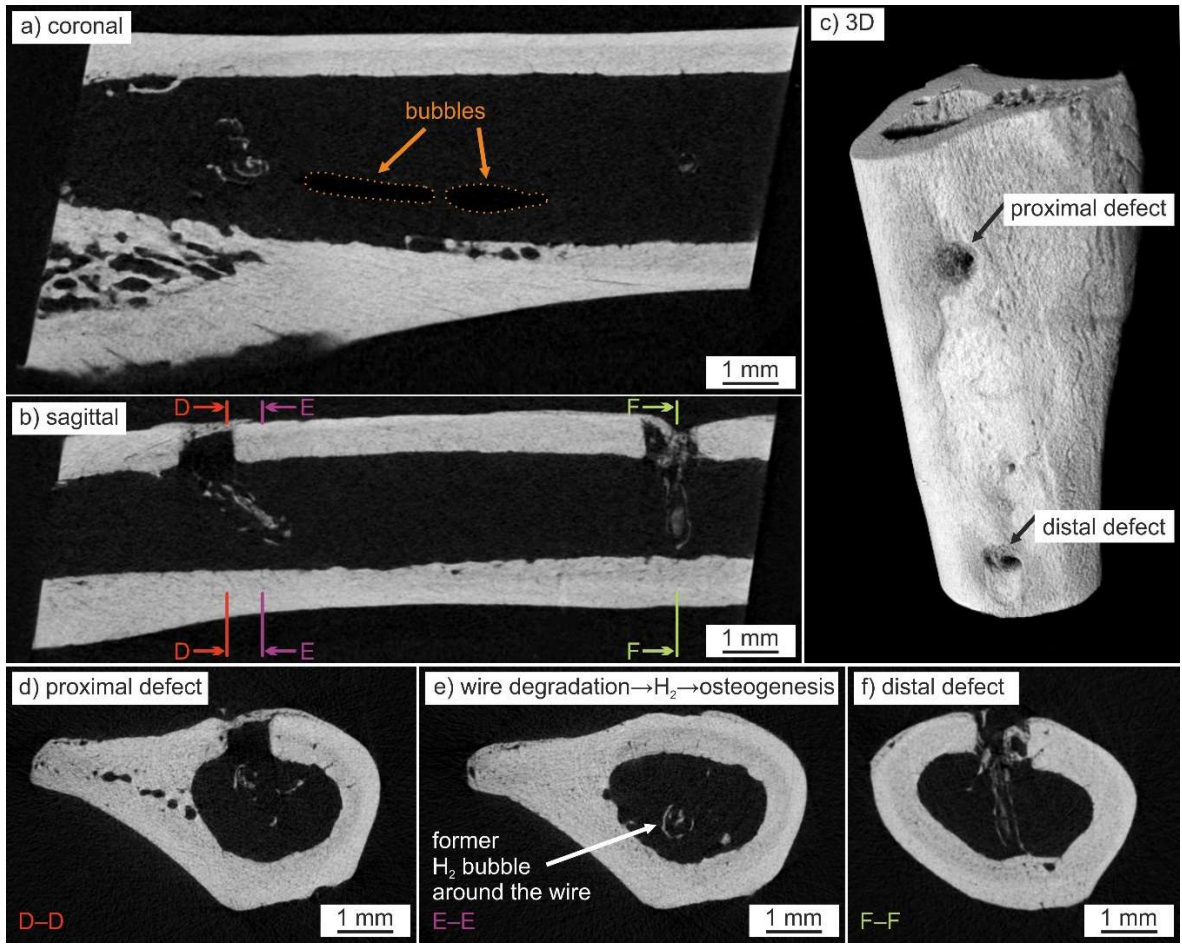

Fig. 7 μCT overview of rat femur with polymer-coated Mg-0.4Zn wire after 1 month of healing (stage 2), a) coronal cut with large bubbles, b) sagittal cut showing wire degradation products and new bone formation, c) 3D overview, d) off-axis cut of the proximal defect, e) site near proximal defect with ring-like osteogenesis, f) distal defect.

The blind histopathological evaluation was conducted on both, transaxial sections of the transcortical defect sites, and the areas between them. The cuts were evaluated in both stages (1 and 2), using a semiquantitative scoring system according to Reifenrath et al. [47]. The results concerning the individual test animals and histological slides are included in Supplement 2. In stage 1, no giant cells that arise as a reaction to the foreign body were observed in any of the histological slides (3-4 slides per animal). Macrophages were detected for all test animals. Moreover, the results suggest a more common occurrence of macrophages for the CP-Mg and uncoated Mg-0.4Zn wires (2 of 3 slides each). In contrast, the test animal with a wire that remained (see Fig. 6b), showed macrophages only in 1 of 4 investigated slides. Furthermore, several neutrophils and periostitis were observed for a CP-Mg wire animal (2 of 3 slides), indicating the presence of inflammation. This, in addition to the poor mechanical properties of the pure Mg, resulted in the discontinuation of trials with CP-Mg for stage 2.

The most apparent results from stage 2 of the histopathological evaluation are summarized in Tab. 4b. When more test animals (4 per condition) and histological slides were examined, slight differences between uncoated (17 slides) and polymer-coated (15 slides) wires emerged. The trend of the negligible number of giant cells still holds for stage 2. As for the macrophages, fewer of them were observed for polymer-coated samples, which could be caused by a slower degradation rate of these wires, as is also suggested by the µCT results (Tab. 4a). The results from the scoring of periosteal and endosteal remodeling also suggest slower degradation of the polymer-coated wires. In general, apart from the trivial difference in terms of the transcortical hole closure, no major histopathological difference is observable between the test animal still containing a wire (polymer-coated wire, stage 1) and the four test animals in stage 2 with the same material which achieved full degradation. This suggests good bone healing properties of PLA/PCL-coated Mg-0.4Zn implants. No direct evidence of the presence of hydrogen bubbles at the time of histopathological evaluation was observed on any of the total 42 investigated histological slides, yet on µCT, they were identified in all stage 1 test animals and in 1 of 8 of the stage 2 animals. This could be caused by the tendency of the largest of the hydrogen bubbles to migrate out of the cross sections of interest of histopathological evaluation and therefore µCT should always be used to determine the hydrogen formation.

## 4   Discussion

### 4.1   Benefits of low-alloyed magnesium for medical devices

The results in this work suggest that the maximum achievable tensile properties of 300 µm Mg-0.4Zn wires reach promising values when matched to the other magnesium wires of similar diameter that are considered for medical devices. A brief overview of some of the current advances in the field of similar thin wires is shown in Tab. 5. Most of the low-alloyed wires (Mg-0.4Zn, ZX10, Mg-0.8Ca) and some other promising wires (LZ21) reach comparable ultimate tensile strength and a ductility value that should be usable for some biomedical implants [22,23,48]. The higher yield strength of 300 µm Mg-0.4Zn wires, produced in this work, could be attributed to the extrusion texture and the sharply planar equilibrium grain boundaries with Zn segregation, as shown in Fig. 2. This grain boundary Zn segregation, although a candidate for a more detailed investigation, could be the reason behind the yield phenomenon. Also, solute-segregated stacking faults that could also cause the beneficial yield strength of these wires are described for Mg-Zn-Y alloy by Drozdenko et al. [49]. As the initial ingot for the Mg-0.4Zn wire was CP Mg containing MgO particles [20], the relatively large deviations for this wire could be improved by a more careful material quality control. When the more heavily alloyed wires in Table 5, that still reach competitive ductility, are considered (ZEK100, WE43B, Mg-4Li-0.5Y), the increase in the yield strength and ultimate tensile strength is evident [28,50]. For these wires, this is caused by the general phenomena occurring in Mg alloys: supersaturation, precipitation, and dispersion of fine precipitates, resulting in lower grain size after hot extrusion. However, when the rapid loss of ductility for *in vitro* degradation is considered (see Tab. 3), the lower initial ductility of highly alloyed Mg wires could become problematic soon after the implantation (although it might be partially attributed to the low wire diameter and the different test gauge lengths in Tab. 5). Therefore, it is necessary to select a wire alloy according to the final biomedical device considered.

Long-term biocompatibility is an important advantage of Mg wires alloyed with low concentrations of clearly biogenic elements that are present in the body in large (Ca) and relatively large (Zn) quantities. As the patients that would benefit the most from the resorbable orthopedic implants are children, great care needs to be taken when addressing the biocompatibility and long-term toxicity of various alloying elements.

| alloy | Dia. (μm) | E/D | $\sigma_{02}$ (MPa) | $\sigma_{UTS}$ (MPa) | $A_t$ (%) | Ref. |
|---|---|---|---|---|---|---|
| Mg-0.4Zn | 300 | E | 223±6 | 248±4 | 22.0±2.3 | This work |
| LZ21 | 300 | D | 189 | 245 | 15 | [48] |
| ZX10 | 400 | E | 154±2 | 217±1 | 34.3±1.2 | [22] |
| ZX10 | 350 | D | 170±2 | 235±1 | 19.1±1.9 | [22] |
| ZX10 | 280 | D | 160±2 | 222±3 | 18.4±4.3 | [22] |
| Mg-0.8Ca | 274 | D | ~ 160 | 235 | 9.4 | [23] |
| ZEK100 | 274 | D | ~ 370 | 458 | 10.6 | [23] |
| WE43B | 127 | D | 397±1 | 412±1 | 10.9±0.2 | [28] |
| Mg-4Li-0.5Y | 125 | D | 271±7 | 322±1 | 11.8±1.0 | [50] |

Tab. 5 Comparison of the recent Mg-based thin wires for biomedical applications, Dia. – final wire diameter, E/D – extruded/drawn condition, $\sigma_{02}$ – 0.2% offset yield strength, $\sigma_{UTS}$ – maximum engineering stress, $A_t$ – total elongation to failure. Approximate values marked with ~ were roughly deduced from the presented graphs.

Aluminum notwithstanding, to some extent problematic in this regard seems to be lithium. Although Li-containing wires have promising properties [48,50], it is well-documented that Li is prescribed as a mood stabilizer and possesses many neuropsychiatric attributes even when low- and micro-doses are administered [51]. It also seems that Li doses affect (although positively) healthy patients [52]. Since all the mass of the resorbable implants is intended to be processed by the body eventually, prescribing Li-containing implants to children, infants, and pregnant persons should be limited, as there is none nor likely will be any basis for studying the long-term effects of mood-affecting elements on such healthy patients. These concerns should be to some extent kept in mind for most rare earth elements [53] for which there are up to date no specific studies that address their long-term effects on humans, especially children, and infants. This should promote the research of low-alloyed Mg-based resorbable implants, such as polymer-coated Mg-0.4Zn wires, for selected patient categories. However, the lower achievable mechanical properties, that could be acceptable in pediatric implantology, will result in difficult regulatory pathways for the development of such medical devices. Further research of grain boundary engineering and Zn segregation of low-alloyed Mg is therefore desirable.

## 4.2 Surface quality, Mg purity and Zn addition alloying

The results of this work suggest that the surface quality is crucial, and the low Zn alloying is more efficient for corrosion control than ultra-high purity magnesium without alloying. It is well known that the surface quality after the extrusion may be difficult to control (Fig. 1a) [21], [54]. Section 3.2. implies that an efficient cleaning for these specific wires is the Nital pickling or nitric acid pickling [30] as it does not leave any residuals on the wire surface when washed with ethanol and is able to remove any required layer of material that was damaged during the extrusion process. Cleaning with other agents such as mechanical swabbing with a polyurethane sponge using 10 % nitric acid, and immersion in commercial alkaline cleaning solutions intended for magnesium or phosphoric acid of various concentrations, did not produce satisfactory results in our trials.

Fig. 8a demonstrates the effect of the surface quality (acidic pickling) and Mg purity on the corrosion rate. Here, the normalized weight change depicts the sum of the corrosion and nucleation of new corrosion products. A CP Mg wire loses weight immediately, signifying that the corrosion on a compact wire is faster for this wire composition than the nucleation of corrosion products. Once the wire starts segmenting, the surface area increases and the redeposition induces accelerated nucleation of the corrosion products. When ultra-pure Mg with a clean surface is considered, a very slow increase in the weight suggests balanced changes from Mg to corrosion products. After a period (~15 days), several large, localized corrosion products are usually developed (similar to Fig. 4b). These may not be able to hold on to the wire during extraction and drying. Even though some of the Nital-pickled ultra-pure Mg wires survived 2 months *in vitro*, the ability of these thin wires to survive the first month *in vivo* was observed to be false in the limited scope of this study in this regard. This is controlled by the probability of a large hydrogen bubble cluster that initiates accelerated localized corrosion. Moreover, the performance of the ultra-pure Mg is rather sensitive to the surface conditions and the acidic-pickling procedure quality, as can be deduced from hydrogen bubbles in Fig. 5a. However, the superiority of the ultra-pure Mg over CP Mg is clearly depicted. For the as-extruded (acetone-cleaned) wires, the analogous scheme where the loose corrosion products nucleate and are not able to hold on to the wire initiates within the first few hours. Afterward, the large surface area and the redeposition stimulated nucleation provides a quick way to wire segmentation.

The main disadvantage in nitric acid-based etchants is the fact that the results are rather dependent on the precise concentration of the solution and the potency of the etchant is changing in time. Therefore, surface quality needs to be checked for each resorbable Mg product. An example is given for Mg-0.4Zn wire in Fig. 8, where Fig. 8b depicts inadequate surface quality due to low Nital concentration and low pickling time. Microcracks and residuals from the extrusion, present in Fig. 8b, are not detectable in Fig. 8c.

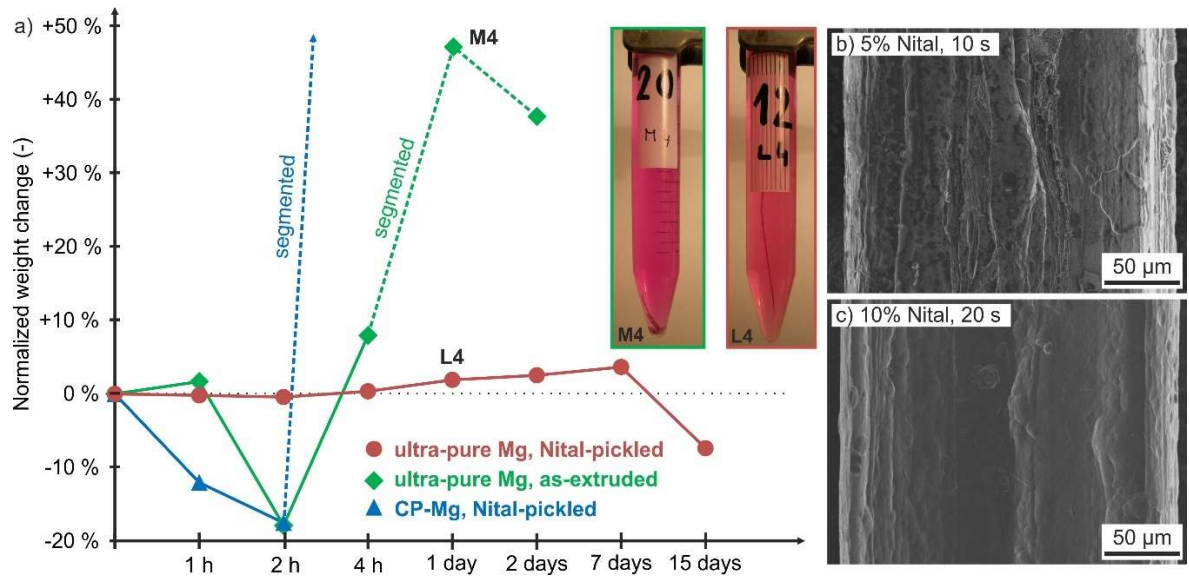

Fig. 8 a) Graph depicting normalized weight change of a dried uncoated Mg wire with respect to material purity and surface condition in αMEM, b) and c) SEM image of Mg-0.4Zn wire surface after 10 s in 5% Nital and 20 s in 10% Nital, respectively.

The conclusions of Jin et al. [55] are in accordance with ours, namely that low-alloyed Mg-Zn alloys seem to be promising for biomaterials due to the low (or no) content of precipitates. The ability of the low-alloyed Mg-Zn wires to accommodate a large or complete fraction of Zn at the grain boundaries (see Fig. 2) is known, and Mg-Zn wires micro-alloyed with Ca could be the logical next step in this direction [56]. The absence of intermetallic MgZn phases is desirable since the Volta potential between Mg and Mg-Zn is rather high and the absence of MgZn binary phases should suppress micro-galvanic corrosion [57]. This is supported by our results (see Fig. 3 and Fig. 5) showing that even low Zn concentration (0.15 wt.%) can improve the *in vitro* corrosion performance of thin Mg wires. However, further research is needed with respect to the corrosion resistance of the Zn-segregated grain boundaries in these wires and other Mg-Zn biomedical devices.

### 4.3 Protective copolymer coating

Several reasons exist for the usage of protective coatings for Mg-based implants, namely, inhibition of corrosion rate, protection of corrosion localization, and surface functionalization [58,59]. For medical devices that need to endure large plastic deformation, e.g., cerclage wires, various polymer coatings can be used. It was shown that these can decrease the initial degradation rate of Mg-based implants [26]. However, the results of 1-month long tests *in vivo*, discussed in Section 3.4, suggest that simple co-polymer coatings used in a metabolically rather active implantation site of the selected animal model do not yield a decrease of *in vivo* corrosion rate for 1-month healing significant enough to prevent wire degradation. These results are affected by the small thickness of Mg-Zn wires and the choice of the implantation site, and also by the selection and thickness of the co-polymer layer.

The corrosion behavior of the copolymer used in our experiments can be considered in terms of its constituents polylactic acid (PLA) and polycaprolactone (PCL), which are chemically biodegradable polyesters. As such, hydrolysis is the main degradation mechanism *in vitro* and *in vivo*. Various studies show that the degradation of these polymers in aqueous environment occurs in the whole bulk [60–62]. This is caused by the fact that water molecules can diffuse into the polymer bulk which limits the protective capabilities as the solution can penetrate up to the metallic Mg-0.4Zn core. The rate of hydrolysis of PLA strongly depends on pH and temperature which leads to different degradation rates within different testing conditions. Newly formed carboxylic groups can increase the speed of hydrolysis due to the local decrease of pH, a mechanism that competes with the other effects that alter pH value in the vicinity of a degrading implant *in vivo* [17]. The monomer and small oligomer units cleaved within the bulk slowly diffuse outside of a material which leads to the formation of cavities. However, the mobility of monomers and oligomers is much slower compared to the water diffusion within PLA. Besides this process, the degradation of PLA is sped up by local inflammation and by enzymes produced by immune cells such as acid phosphatase or lactate dehydrogenase [62]. The degradation of PCL, and its copolymers, follow a pattern similar to PLA [60]. The degradation gradually decreases the molecular weight of the molecules and upon reaching 10 kDa or smaller, PCL mechanical stability becomes significantly compromised. Furthermore, molecules of 10 kDa or less can undergo intracellular degradation. It is worth noting that the crystalline phase of PCL is more resilient to hydrolysis compared to the amorphous one [61]. When the corrosion rate of thin Mg-based biomedical devices is considered, more complex polymer or composite coatings should be developed [63]. Polymer coatings based solely on PLA and similar polyesters should be mainly used for thin Mg-based implants where the primary goal is surface functionalization. However, there are novel studies, where pure PCL coating has led to a promising decrease in the *in vivo* corrosion rate, for example [48]. Further research in this regard is therefore justifiable, although the effect of the implantation site should be considered. Namely, when addressing the *in vitro-in vivo* correlation and polymer protection effectiveness, subcutaneous *in vivo* implantation sites could result in different degradation rates than artificially induced bone defect sites [64]. The results of this work suggest the importance of full utilization of the metallurgical aspects of alloy design, as protective coatings are rarely a universal solution for corrosion rate control of Mg alloys.

### 4.4 Mg-induced osteogenesis and hydrogen-tissue interactions

It is well known that Mg and Zn locally affect bone mineralization and osteogenesis, being divalent metal elements found naturally in the human body [65]. Yet, there exist concerns regarding the level of local $Mg^{2+}$ concentration that should still be beneficial for bone mineralization [66]. The *in vitro* and *in vivo* results in this work suggest that for small implants, binary Mg-0.4Zn promotes osteogenesis substantially. Moreover, low alloying with Zn seems to provide competitive mechanical properties due to various positive effects (Tab. 5), stabilize corrosion performance, and even slightly increase the *in vitro* performance on MG-63 cells (Fig. 5). In terms of *in vivo*, our results (Tab. 4, Supplement 2) are in general accordance with Li et al. [67], namely in the fact that Mg-Zn material shows excellent histocompatibility. The protective layer composed of PLA/PCL copolymer did not affect the histocompatibility negatively and was able to slow down the corrosion of the implants. However, the results of this work suggest that the polymer layer used here is not an efficient tool for suppressing the effect of alternate surface quality of the wire concerning the *in vivo* degradation performance. The vastly different outcomes for stages 1 and 2 in this study imply that great care needs to be taken in future investigations of Mg-based implants, especially in a case when a quantitative correlation of *in vivo* degradation rates between different studies is performed.

The rat femur model, used in this work, provides a valuable implant site for various studies [67,68]. Fig. 9 shows a transaxial cut in the distal defect in an animal implanted with polymer-coated Mg-0.4Zn wire. Local osteogenesis in the medullary cavity is depicted by various techniques in the vicinity of the original wire position. For the polymer-coated wires in this study (generally being in the earlier stages of the bone resorption and remodeling process), the wire degradation products in the intramedullary cavity (apparent from increased Mg content in Fig. 9f) are surrounded by a layer of non-mineralized tissue (up to ~200 μm) and then a thin layer of bone tissue which is not that rich in Mg.

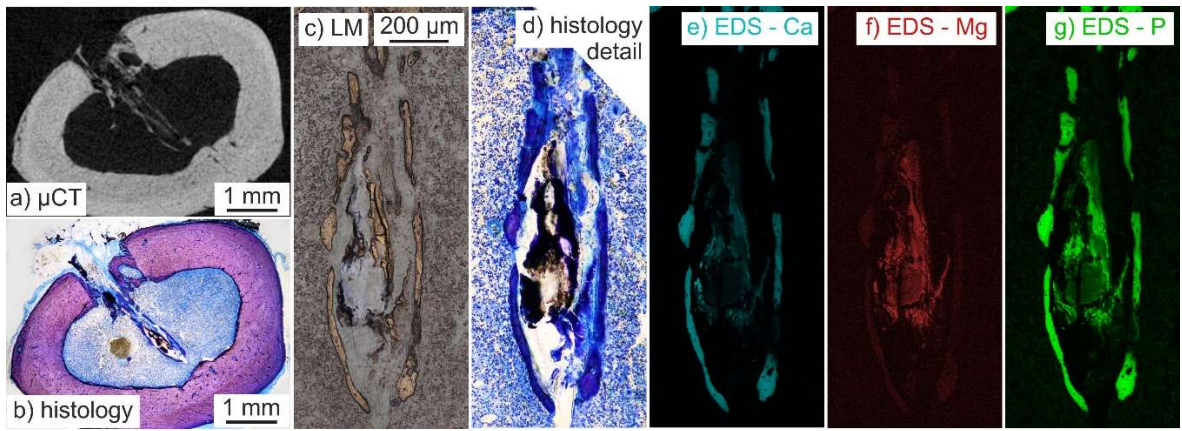

Fig. 9 Distal defect with remnants of degraded polymer-coated Mg-0.4Zn wire and a convex shape of a bone tissue envelope a) μCT cut, b) histological slide, c) LM image of Mg-induced osteogenesis, d) histological detail, e-g) SEM-EDS maps.

From the convex (Fig. 9c-g) or spherical shapes (Fig. 7e) of the surrounding bone tissue, commonly observed in the evaluated samples, a question arises, whether the corrosion of such small implants in the medullary cavity does not follow a route, vaguely described as i) Mg degradation products (such as Mg-substituted calcium phosphates) continually replace Mg implant, ii) a hydrogen envelope is formed, iii) new bone, not rich in Mg, is formed on the tissue/H$_2$ interface, iv) hydrogen envelope is slowly replaced by fibrous or non-mineralized tissue, v) remaining bone tissue is slowly resorbed and a normal appearance of the medullary cavity is restored. Not enough data have been gathered to support this sequence without a doubt, therefore, it is stated here as a motivation for further studies.

The formation of hydrogen in the vicinity of Mg-based implant and the related peri-implant gas accumulation is a complex problem [69] and larger animal models need to be used as well [70] before a full understanding of the hydrogen-tissue interactions in the healing processes can be reached. Nevertheless, Mg-based cerclage wires and strands will find their utilization in various bone-related biomedical applications, surpassing the currently used non-degradable fixation wires and sutures and polymer-based biodegradable materials.

## 5 Conclusions

Thin biodegradable wires with a diameter of 250 µm or 300 µm were produced by a single-step hot direct extrusion process from ultra-pure Mg, commercially pure Mg, Mg-0.15Zn, Mg-0.4Zn and Mg-1Zn alloys. Selected wires were coated with copolymer of L-lactide and ε-caprolactone in a 70/30 molar ratio. The resulting Mg-0.4Zn wires provide an interesting possibility to be employed as cerclage wires and other bone fixation devices, especially in coated and stranded conditions for higher reliability. The individual outcomes of this study can be summarized as follows:

1. Extruded Mg-0.4Zn wires offer exceptional mechanical properties by reaching small grain size, absence of MgZn binary phases, prominent grain boundary Zn segregation, and elevated yield strength due to the Zn-governed yield phenomena.
2. Limited Zn alloying (0.4 wt.%) of commercially pure Mg is superior in terms of corrosion control to the ultra-pure Mg (Mg alloying with Zn > Mg purification).
3. *In vitro* analysis of cell adhesion and proliferation on MG-63 cells of ultra-pure Mg and Mg-0.15Zn showed comparable cell behavior to the control after 6 days, with Mg-0.15Zn being marginally better.
4. Performance of small Mg implants is sensitive to the surface quality, both *in vitro* and *in vivo*. Acidic pickling with nitric acid can remove surface contamination and damage. Used PLA/PCL copolymer coating is not able to fully counter poor initial Mg surface quality. However, the copolymer slightly decreases the *in vivo* degradation rate.
5. Both uncoated and copolymer-coated Mg-0.4Zn wires promoted *in vivo* osteogenesis in the vicinity of the implants.
6. Preferential bone tissue formation on the hydrogen-tissue interface is suspected as an important part of the intramedullary Mg-0.4Zn wire degradation mechanism.

**CRediT author statement**

**K. Tesař:** Conceptualization, Methodology, Validation, Investigation, Resources, Data Curation, Writing - Original Draft, Writing - Review & Editing, Visualization, Supervision, Project administration, Funding acquisition, **J. Luňáčková:** Methodology, Validation, Formal analysis, Investigation, Data Curation, Writing - Original Draft, Writing - Review & Editing, Visualization, **M. Jex:** Formal analysis, Writing - Original Draft, Writing - Review & Editing, **M. Žaloudková:** Methodology, Investigation, Writing - Review & Editing, Visualization, **R. Vrbová:** Methodology, Formal analysis, Investigation, Writing - Review & Editing, Visualization **M. Bartoš:** Methodology, Formal analysis, Investigation, Writing - Original Draft, Writing - Review & Editing, Visualization **P. Klein:** Methodology, Investigation, Resources, Writing - Original Draft, Writing - Review & Editing, **L. Vištejnová:** Methodology, Investigation, Writing - Review & Editing, **J. Dušková:** Methodology, Formal analysis, Investigation, **E. Filová:** Methodology, Formal analysis, Investigation, Data Curation, Writing - Review & Editing, **Z. Sucharda:** Investigation, Resources, **M. Steinerová:** Investigation, Formal analysis, Writing - Review & Editing, **S. Habr:** Methodology, Resources, **K. Balík:** Resources, Funding acquisition, **A. Singh:** Methodology, Validation, Investigation, Writing - Original Draft, Writing - Review & Editing.

**Acknowledgments**

K.T. acknowledges the project FerrMion of the Ministry of Education, Youth and Sports, Czech Republic, co-funded by the European Union (CZ.02.01.01/00/22_008/0004591). The authors acknowledge the support of The Charles University Grant Agency in the frame of the project No. 121724 and the project Cooperatio No. 207030 Dental Medicine/LF1 of the Charles University. M.J. received financial support from the Ministry of Education, Youth and Sport of the Czech Republic under the grant No. RVO 14000. J.D. was supported by the Ministry of Health of the Czech Republic - RVO project VFN64165. M.Z., K.B. and Z.S. acknowledge the support of the project GAMA 2 of the Technology Agency of the Czech Republic No. TP01010055. E.F. and M.S. acknowledge the project of the Czech Academy of Sciences, Czech Republic (Praemium Academiae grant No. AP2202). P.K. and L.V. acknowledge the support of the Ministry of Health of the Czech Republic, grant project No. NU20-08-00150. The authors would like sincerely to thank T. Jurkovičová from Laboratory of Analytical Chemistry of FZU for the AAS and A. Jančová, J. Maňák, J. Liška, M. Křížková, Z. Weiss, P. Nekoksa, J. Lukšíček, M. Plisová, J. Ryjáček and J. Havlíček for technical assistance and the company Crown Metals CZ, s.r.o. for the continued support in the frame of Mg-based materials donations for basic research.

**Declaration of interest**

Karel Tesař and Karel Balík declare that Czech Technical University in Prague, Institute of Physics of the Czech Academy of Sciences and Institute of Rock Structure and Mechanics of the Czech Academy of Sciences have jointly filed a national patent application (Pending, PV 2023-459) regarding the Nital-pickling procedure for Mg alloys, used in this work. Others declare no conflict of interest.